\documentclass[preprint2]{aastex}
\begin{document}
\title{Membership Determination of Open Cluster M48 Based on BATC Thirteen-Band
Photometry}
\author{Zhen-Yu Wu, Xu Zhou, Jun Ma, Zhao-Ji Jiang, Jian-Sheng Chen}
\affil{National Astronomical Observatories, Chinese Academy of
Sciences, 20A Datun Road, Beijing 100012, China}
\email{zywu@bac.pku.edu.cn}
\begin{abstract}
Beijing-Arizona-Taiwan-Connecticut (BATC) multi-band photometric data in the
field of open cluster M48 are used to determine its membership. By comparing
observed spectral energy distributions (SEDs) of stars with theoretical ones,
membership probabilities of 750 stars with limiting magnitude of 15.0 in BATC
$c$ band ($\lambda_{eff}=4194$\,\AA) are determined. 323 stars with membership
probabilities higher than 30\% are considered as candidate members of M48.
Comparing membership probabilities of 229 common stars obtained by the present
method and the proper-motion based methods, a 80\% agreement among these methods
is obtained.

\end{abstract}
\section{Introduction}
Open clusters (OCs) have long been recognized as important tools in the study of
the Galactic disk. They have been used to determine spiral arm structure, to map
the rotation curve of the Galaxy, to investigate the mechanisms of star
formation, to constrain the initial luminosity and mass functions, and to
define disk abundance gradients and age-metallicity
relationship \citep{fr95,fr02,ch03,sal04,von05,bo06}.

The first step to determine the physical parameters of an open cluster is to
select probable members in the vicinity of this cluster. Omitting the parameter
of position which provides us with the first clue as to the existence or
nonexistence of a cluster, there are two types of independent methods by which
to establish cluster membership: photometric and kinematic \citep{ca90}.

When kinematic data are available, it is accepted that the membership
probabilities obtained from the analysis of proper motions or radial velocities
are more reliable. Unfortunately, few clusters have a large, homogeneous radial
velocity data to permit a detailed analysis of the entire cluster. It is well
known that the proper motion analysis is at present the most valuable criterion
to establish membership probabilities in OCs \citep{sl77,ca85}.

The first attempt to statistically determine the membership of an open cluster
based on proper motion data was made by \citet{va58}. They point out that the
cluster and field probability density functions can be modeled as bivariate
Gaussian distributions: a circular normal distribution for the cluster
population,  and an elliptic normal distribution for the field population. A
maximum likelihood principle was developed to obtain the distribution parameters
of clusters and membership probabilities of individual stars \citep{san71}. The
parametric Vasilevskis-Sanders method has been frequently used to derive the
membership of star clusters \citep{wu02a, wu02b}. However, the hypotheses for
cluster and field stars distribution in the parametric Vasilevskis-Sanders
method are not always true. Even if the hypotheses are realistic for some
clusters, the parametric method will fails when the cluster member-to-field star
ratio is small \citep{ca90}. This method does not work in the case of
significant internal motion in a cluster or its rotation \citep{sl77,ja06}.

In order to overcome some difficulties that arise from the parametric
Vasilevskis-Sanders method, \citet{ca90} developed a non-parametric approach to
the membership problem. The key of their method is to perform an empirical
determination of the probability density functions, without relying on any
previous assumption about their profiles \citep{ga98}. The non-parametric method
has been used in recent years to determine membership of several OCs
\citep{ga98, ba04, ba05}. More recently, \citet{ja06} presented a new method
that enlarges the statistical distance between the cluster members and field
stars by revealing the group of stars with the least relative velocities without
any assumptions about the distribution of field stars.

Although proper-motion based methods have been considered to be the most
reliable for cluster -- field segregation, however, these methods need high
precision proper motion data. Several decades or even more than one hundred
years are needed to obtain these high precision proper motion data. So,
membership determination based on proper motions is very time-consuming. On the
other hand, most of proper motions of stars are derived from plate data, high
precision results can obtained only for bright stars.

Color-magnitude diagrams (CMDs) or color-color diagrams are used to derive the
fundamental parameters of clusters, and in general, are also used to determine
the membership of cluster. If a field of cluster and, a comparison field which
only including field stars, are both obtained for study, the CMD or color-color
diagram of comparison-field stars are subtracted from that of field including
the cluster, the resulting difference diagram will show most of members of this
cluster \citep{mei00,von02}. However, the method obviously requires that the
foreground and background of both the cluster and the comparison field, are
essentially identical, but above conditions are not satisfied due to the
non-uniformity of the background \citep{ba83}. On the other hand, if no
comparison field is obtained, theoretical isochrones are used to match the
observed star sequence, and stars residing in the neighborhood of the
best-fitting isochrone are considered as members of this cluster. At the mean
time, the fundamental parameters of this cluster such as age, metallicity,
distance, and reddening can be derived \citep{jef01}. In general, it can't be
ensured reasonably to derive the four fundamental physical parameters just using
observational results in two or three bands without knowing any one of those
parameters in advance. So, it is much more difficult to determine the members of
cluster just based only on CMDs or color-color diagrams without knowing any
physical parameters of cluster. More recently, using $UBVRI$ and
$uvby\textrm{H}\beta$ photometric data, \citet{sar99} and \citet{twa00} present
new method to distinguish cluster and field stars based only on observed star
sequences in CMDs or color-color diagrams. \citet[and references therein]{kal04}
present the synthetic CMD method to derived the fundamental parameters of
cluster.

The BATC photometric system including 15 bands provides a sort of
\textsl{low-resolution spectroscopy} that defines the spectral energy
distributions (SEDs) of each star. If the membership are known with the
proper-motion based method, fitting the observed SEDs of cluster member stars
with theoretical models has an advantage that it has more observational data
than the number of free parameters to be solved.  Fitting the observed SEDs with
theoretical ones can also be used to derive both the membership and fundamental
parameters of a cluster at the same time.

In this paper, we develop a method based on fitting SEDs of stars in a field of
star cluster with theoretical ones to determine the membership and fundamental
parameters of this cluster at the same time. Comparing with traditional
photometric membership determination methods based on CMDs or color-color
diagrams, the advantage of the present method is that no comparison field is
needed and the fundamental parameters of cluster can also be derived more
reliable. Moreover, this method can present the membership probabilities of
stars. Comparing proper-motion based methods, the observational data used by the
present method can be obtained within one month, no any assumptions about the
distribution profiles of cluster and field stars are needed, and membership
probabilities of stars with more fainter magnitude can be obtained. We apply
this method to open cluster M48 with 13 bands of BATC photometric data. Section
\ref{data} describes the proper motion and photometric data as well as
theoretical model used for applying the present method. The details of our
present method are presented in Section \ref{method}. We apply our present
method to open cluster M48 and compare our results with those derived by
proper-motion based methods in Section \ref{disc}. Finally, a summary is
presented in Section \ref{conc}.

\section{Data and Theoretical Model\label{data}}
\subsection{Data Used}
The open cluster M48, also known as NGC 2548, is quite a conspicuous object and
is visible to the naked eye under good weather conditions. Two elements make us
to choose this cluster as the first object to apply our present method to
determine its membership. This cluster has been observed and calibrated in 13
bands of the BATC photometric system in our previous paper \citep[hereafter
Paper I]{wu05}. The BATC photometric system consists of 15 filters with
bandwidths of 150 -- 350\,\AA  \space covering the wavelength range 3300 --
10000\,\AA, and avoiding strong and variable sky emission lines \citep{fan}. A
60/90 cm f/3 Schmidt telescope was used, with a Ford Aerospace $2048\times2048$
CCD camera at its main focus. The field of view of the CCD is $58\arcmin \times
58\arcmin$, with a scale of $1\farcs7$ pixel$^{-1}$. The definition of magnitude
for the BATC system is in the AB$_{\nu}$ system, which is a monochromatic flux
system introduced by \citet{og}. The details of data reduction and photometric
results are described in Paper I.

The membership probabilities of stars in the field of M48 have been determined
by \citet{wu02b} with parametric Vasilevskis-Sanders method as well as by
\citet{ba05} with non-parametric method. Proper motions of 501 stars within a
$1\fdg6 \times 1\fdg6$ area in the region of M48 are given by \citet{wu02b}. Ten
plates of this cluster are used to derive proper motions. The oldest plate was
taken in 1916, and the newest ones in 1998. The RMS errors on proper motions for
more than 90\% of stars are $\varepsilon_{\mu_{\alpha}\textrm{cos}\delta}=0.92$
mas yr$^{-1}$, $\varepsilon_{\mu_{\delta}}=0.68$ mas yr$^{-1}$.  By applying the
parametric Vasilevskis-Sanders method and using a 9-parametric Gaussian model
for the frequency function, \citet{wu02b} concluded that stars with membership
probabilities higher than 70\% are members. \citet{ba05} re-analyzed the proper
motion data of \citet{wu02b} with a non-parametric method \citep{ga98,ba04},
they concluded that stars with membership probabilities as high as 82\% based on
non-parametric method or with membership probabilities as high as 92\% based on
parametric method, are the most probable cluster members.

So, we chose M48 as the first object to apply our present method based on BATC
multi-band photometric data and to compare our membership determinations to
those derived from proper motion data.

\subsection{Theoretical Model}
Padova stellar evolutionary models (Padova 2000) \citep[and references
therein]{gir00,gir02} are used in our present method. Padova 2000 models
present a large grid of stellar evolutionary tracks and isochrones, which are
suitable to modeling star clusters. The isochrones are presented for the initial
chemical compositions: $Z=0.0004, Z= 0.001, Z=0.004, Z=0.008, Z=0.019$ (solar
composition), $Z=0.03, Z=0.04$, and $Z=0.07$. These models are computed with
updated opacities and equation of state, and a moderate amount of convective
overshoot. The range of initial masses goes from 0.15 $M_{\odot}$ to 7
$M_{\odot}$, and the evolutionary phases extend from the zero age main sequence
to either the thermally pulsing AGB or carbon ignition. They also present an
additional set of models with solar composition, computed using the classical
Schwarzschild criterion for convective boundaries.

Using the method from \citet{gir02}, the basic output of stellar models -the
surface luminosity $L$ and effective temperature $T_{eff}$ are converted into
the observable quantities, i.e. magnitudes and colors in BATC photometric
system. \citet{gir02} re-write the formalism for converting synthetic stellar
spectra into tables of bolometric corrections. The resulting formulas can be
applied to any photometric system, provided that the zero-points are specified
by means of either ABmag, VEGAmag, or a standard star system. They also assemble
an extended and updated library of stellar intrinsic spectra, which is mostly
based on ``non-overshooting'' ATLAS9 models \citep{ca97}, suitably extended to
both low and high effective temperatures. This offers an excellent coverage of
the parameter space of $T_{eff}$, $\log g$, and [M/H] \footnote{The Padova
stellar evolutionary models in BATC photometric system can be download at
``http://pleiadi.pd.astro.it/isoc\_photsys.02/isoc\_batc/index.html''.}.

\section{Method\label{method}}
In a  star field including a star cluster, the stars can be divided into many
groups which share the same age, distance, metallicity, and reddening. The
astronomical hypotheses in the present method
can be set in the following way that the group including member stars of the
cluster must have the maximum star number in the studied star field and must
share the same distance, age, metallicity and reddening. In our assumption,
common reddening for all member stars in a cluster not always a true one.
However, the reddening maps of \citet{sch} show very small differential
reddening across the
field of M48, hence, the assumption of common reddening for all member stars is
reasonable.

Based on above assumptions, the problem of membership determination of a cluster
can be resolved by finding the star group that includes the maximum star number.
In order to divide stars into different star groups according to their physical
parameters: age, metallicity, distance, and reddening, theoretical stellar
models with various different sets of physical parameters must be used. By
fitting the observed SEDs of stars in the studied field with the theoretical
model with a given set of physical parameters, stars, whose observed SEDs can be
best fitted by the same theoretical ones, are determined as the members of that
star group and the physical parameters of this group are obtained from the
theoretical model. Above SED-fitting processes are repeated for different
theoretical models with different sets of physical parameters. In the end, stars
in the group with the maximum members are considered as the cluster's members of
the studied star filed.

The observed SEDs of a star is determined by intrinsic (mass, age, and
metallicity) and extrinsic (distance and reddening) parameters. In order to
apply our present method, our fitting procedure is separated into two
steps. First, a theoretical isochrone with different stellar masses but with the
same age and metallicity is chosen, then, a distance modulus and reddening
correction are applied to this theoretical isochrone. For $j$th star , a
parameter $S$ can be defined:
\begin{equation}
S_{j}(t,Z,d,E(B-V))=\sum_{i=1}^n\frac{(m_{ij}-M_{i}(t,Z,d,E(B-V))^{2
} } { \sigma_ { ij}^{2}}
\end{equation}
where $M_{i}(t,Z,d,E(B-V))$ is the theoretical magnitude in BATC $i$th band
corrected by distance modulus $d$ and reddening $E(B-V)$, computed from the
chosen theoretical isochrone model with age $t$, metallicity $Z$, and mass m.
The reddening $E(B-V)$ is transformed to each BATC band with the extinction
coefficient derived by \citet{ch00} based on the procedure in Appendix B of
\citet{sch}. $m_{ij}$ and $\sigma_{ij}$ is the observed magnitude and its error
of $j$th star in $i$th band. $n$ is the total number of observed bands for $j$th
star. For $M_{i}$ with different stellar masses, the minimum of $S_{j}$:
$S_{j\textrm{min}}$ can be obtained for the $j$th star with the chosen
theoretical models. For a star sample, we repeated above fitting process for
each star and calculate the $S_{\textrm{min}}$ parameters for them. If the
observed SEDs can match the theoretical ones, the parameter $S_{\textrm{min}}$
should be $\chi^{2}$ distribution with $n-P$ degrees of freedom, where $P$ is
the number of free parameters to be solved. The integral probability at least as
large as $S_{j\textrm{min}}$ in $\chi^{2}$ distribution with $n-P$ degrees of
freedom is taken as the membership probability of $j$th star in the given
cluster with the theoretical physical parametric set. If the membership
probability is large than 30\%, the star $j$ is considered as a member star for
that model with the given set of physical parameters. For each set of physical
parameters, another parameter can be defined:
\begin{equation}
S_{c}(t,Z,d,E(B-V))=\frac{\sum_{j=1}^NS_{j\textrm{min}}}{N}
\end{equation}
where $N$ is the number of member stars.

We repeated the above process for various parametric sets with different
distance modulus, reddening, age and metallicity. A parametric set with the
maximum $N$ and minimum $S_{c}$ is considered as the best-fitting one for this
cluster and the member stars in this cluster are also determined at the same
time.

\section{Result and Discussion\label{disc}}
We apply the present method to open cluster M48, the parameters are chosen as
following: metallicity are taken as $Z=0.004,0.008,0.019,0.03,0.04$, ages
$\log(t)$ from 8.3 to 8.8 with a step of 0.05, distance modulus from 9.0 to 9.7
with a step of 0.01 and reddening $E(B-V)$ from 0.00 to 0.15 with a step of
0.01. 750 stars, whose BATC $c$ ($\lambda_{eff}=4194$\,\AA) magnitude is
brighter than 15.0, are chosen for investigation. The limiting magnitude of 15.0
is 2 magnitude deeper than that used by proper-motion based methods
\citep{wu02b,ba05}.

\begin{figure}
 \epsscale{1.0}\plotone{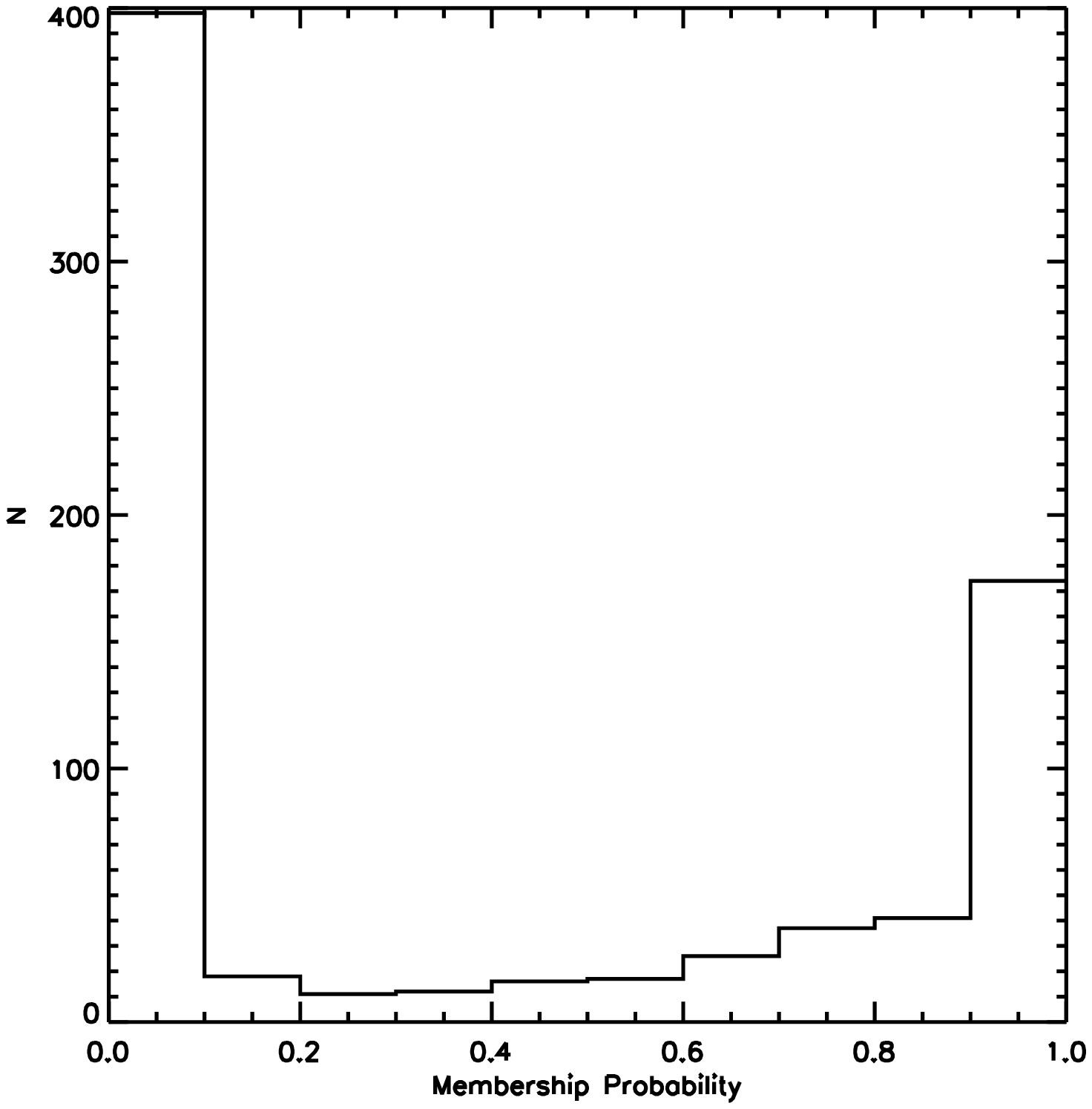}\caption{The histogram of cluster
membership probability of M 48. The present SED method is
used to derive the membership probabilities for 750 stars with
limiting magnitude of 15.0 in BATC $c$ band.\label{hist}}
\end{figure}

In the end, we find a best-fitted set of physical parameters
from theoretical model with an age of $\log(t)=8.65$, a metallicity $Z=0.008$, a
distance modulus 9.05 and $E(B-V)=0.10$ . The derived parameters are consistent
with the previous determinations within the errors \citep{ri04,wu05,ba05}. The
cluster membership probability histogram (Figure \ref{hist}) shows a clear
separation between cluster members and field stars. The number of stars with
membership probabilities higher than 30\% is 323 and considered as members of
M48. The average membership probability of cluster stars is 84\%, giving a
contamination by field stars not larger than 16\%.

There are 229 stars in our sample which are common with that of \citet{wu02b}.
If we consider the most probable member stars are those with membership
probabilities higher than 82\% (non-parametric method) or 92\% (parametric
method) just as \citet{ba05} pointed out, there are 97 stars are most probable
members of M48 based on proper motion data. Among the 97 stars, 81 stars are
considered as members with membership probabilities higher than 30\% in this
paper, another 16 stars are considered as field stars. Among another 132 common
stars which considered as field stars by proper-motion based method, there are
33 stars are considered as members by the present SED method. With these
limiting membership probabilities (82\% for non-parametric or 92\% for
parametric method and 30\% for the present method), we get a 80\% agreement in
the segregation yield by the present method and proper-motion based method. It
should be noted that these most probable member stars determined by both
parametric and non-parametric method based on proper motion data, are used for
comparison. If stars with membership probabilities higher than 70\% are
considered as members of M48 by parametric method of \citet{wu02b}, there is a
86\% agreement between the present method and the parametric method.

\begin{figure}
 \epsscale{1.0}\plotone{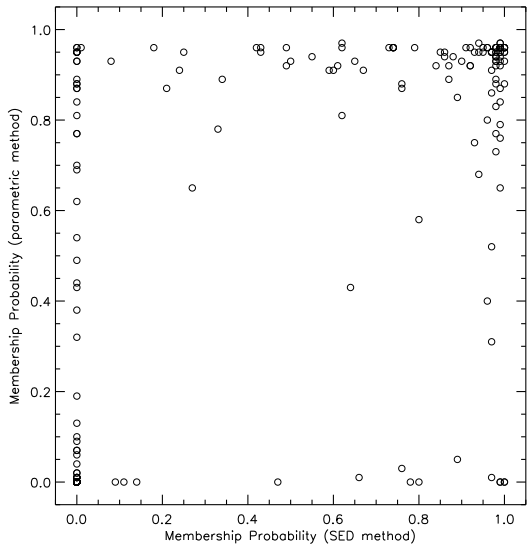}\caption{Comparison of membership probabilities
for common stars between the SED method and the proper-motion based
parametric method.\label{prb}}
\end{figure}

\begin{figure}
\epsscale{1.0}\plotone{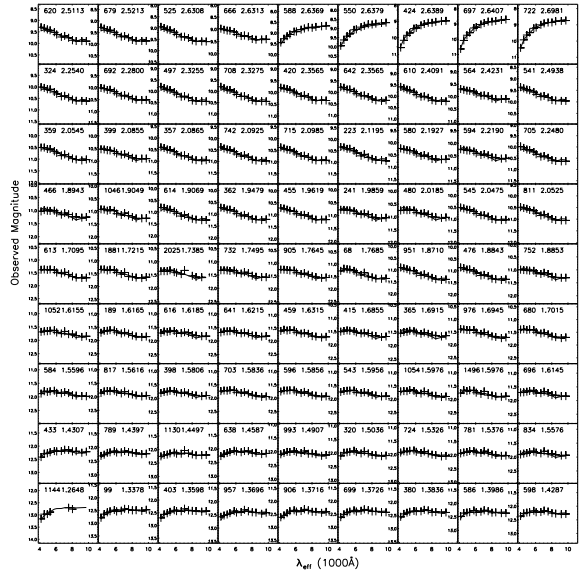}\caption{SEDs of 81 member stars of M48 determined
by present SED method  as well as by parametric  and non-parametric method based
on proper motion data. Symbols '+' represent the observed magnitudes in each
BATC band and the solid lines connect the best-fitting theoretical values. In
each panel, IDs from Paper I and best-fitting mass for each star are also
labeled.\label{f1}}
\end{figure}

In Figure \ref{prb}, we compare membership probabilities determined by the
present method and the proper-motion based parametric method. Although the
statistical nature of the present method and the proper-motion based methods are
different, Figure \ref{prb} shows the consistent results in the high membership
probability region obtained by these two kinds of methods. Figure \ref{f1} shows
SEDs of 81 member stars of M48 determined both by the present SED method
(membership probabilities higher than 30\%) and by proper-motion based methods
(membership probabilities $\geq$ 92\% for parametric method or membership
probabilities $\geq$ 82\% for non-parametric method). Symbols `+' represent the
observed magnitudes in each BATC band and the solid lines connect the
best-fitting theoretical values. In each panel, IDs from Paper I and the
best-fitting mass for each star are labeled. It can be seen from Figure
\ref{f1}, the theoretical SEDs can fit the observed ones very well both for
giant and MS stars. The changes of SED's shapes of stars along with magnitudes
(masses) are also showed clearly in Figure \ref{f1}. Table \ref{tb1} lists these
81 member stars of M48 determined by both the present SED method and the
proper-motion based methods (parametric or non-parametric method). Column 1 is
the identification number from Paper I; Cols. 2 -- 14 are BATC magnitudes and
their errors in 13 bands for each star; Cols. 15, 16, and 17 are the membership
probabilities of the present SED method, \citet{wu02b}, and \citet{ba05},
respectively, for each star; Col. 18 is the identification number in
\citet{wu02b}. Table \ref{tb2} lists the member stars of M48 only determined by
the present SED method. The format is the same as in Table \ref{tb1}. If there
is no identification number in \citet{wu02b} for a star, a `0' is given in Col.
18.

\begin{figure}
\epsscale{1.0}\plotone{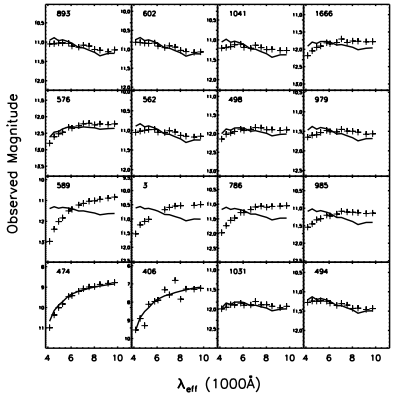}\caption{SEDs of 16 field stars determined by
present SED method but considered as member stars by parametric method and
non-parametric method based on proper motion data. Symbols and line type have
the same meaning as in Figure \ref{f1}. Only IDs from Paper I is labeled in each
panel.\label{f3}}
\end{figure}

\begin{figure}
 \epsscale{1.0}\plotone{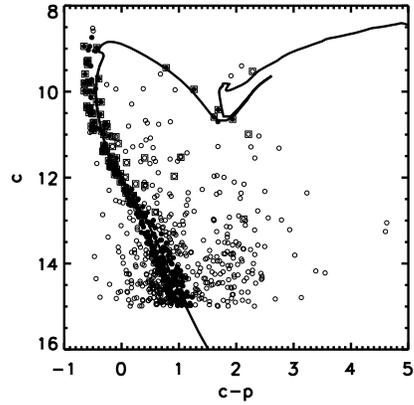}\caption{CMD of M48 with BATC $c$ and $p$ bands.
All stars in our sample with limiting magnitude of 15.0 in BATC $c$ band are
plotted as open circles; member stars determined by present SED method are
plotted as filled circles and member stars determined based on proper motions
are plotted as open squares. A distance modulus of 9.05, reddening $E(B-V)=0.10$
are adopted. A Padova isochrone with an age of $\log(t)=8.65$ and metallicity
$Z=0.008$ is over-plotted in the CMD with a solid line.\label{cmd}}
\end{figure}

In Figure \ref{f3}, we plot SEDs of 16 field stars determined by the present SED
method with membership probabilities less than 30\%, but considered as member
stars by the parametric method (membership probabilities $\geq$ 92\%) or the
non-parametric method (membership probabilities $\geq$ 82\%) based on proper
motion data. Only IDs from Paper I are labeled in each panel. For stars (no.
406, 474, 494, and 1031), the theoretical SEDs can't fit the observed ones due
to large deviation in some bands and these four stars are probably members. For
stars (no. 3, 786, 589 and 576), the observed SEDs of these stars show that they
are probably background red giants. For another stars, the theoretical SEDs
can't fit the observed ones well, and they are probably double stars. In Figure
\ref{cmd}, the CMD of M48 with BATC $c$ and $p$ bands is plotted. All 750 stars
in our sample are plotted as open circles, member stars determined by the
present SED method are plotted as filled circles and member stars determined by
the proper-motion based methods are plotted as open squares. Adopting the
derived distance modulus of 9.05 and reddening $E(B-V)=0.10$, a Padova
theoretical isochrone with an age of $\log(t)=8.65$ and metallicity $Z=0.008$ is
over-plotted in the CMD. From this CMD, it can be seen that most stars
considered as members by \citet{wu02b} and \citet{ba05}, but considered as field
stars by the present method, locate in the range of CMD where $12.5>c>10.7$ and
$c-p>0.2$ and lie at the right of main sequence of this cluster. From the above
analysis based on SEDs of those stars, most of them are probably double stars or
background red giants. Table \ref{tb3} lists these 16 field stars determined by
the present SED method but considered as member stars by the proper-motion based
methods. The format is same as Table \ref{tb1}.

\begin{figure}
 \epsscale{1.0}\plotone{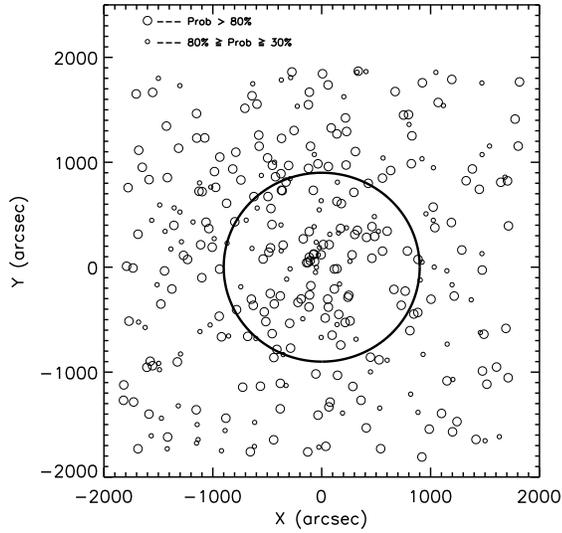}\caption{The spatial distribution of member stars
determined by the present SED method.
The
sizes of symbols represent the different membership probabilities for each
star and labeled on the top. A circle with diameter of 30 arcmin is also plotted.\label{xy}}
\end{figure}

\begin{figure}
 \epsscale{1.0}\plotone{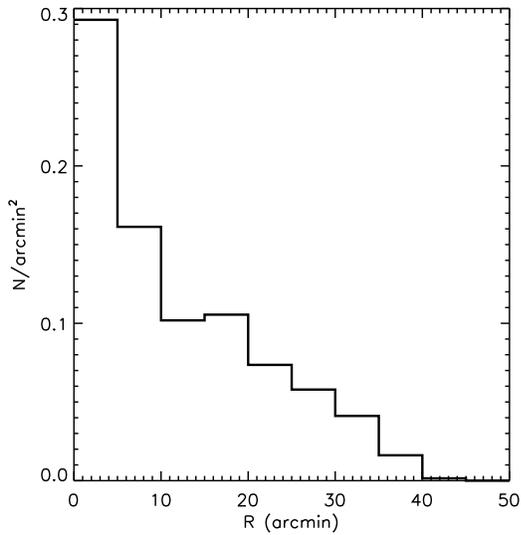}\caption{The density distribution of member stars plotted in Figure \ref{xy}.\label{den}}
\end{figure}

Figure \ref{xy} shows the spatial distribution of member stars with membership
probabilities higher than 30\% determined by the present method. The sizes of
symbols represent the membership probabilities of each star and are labeled on
the top of Figure \ref{xy}. A circle with diameter of 30 arcmin is also plotted
in Figure \ref{xy}. Figure \ref{xy} shows that most of member stars concentrate
near the center of cluster. In Figure \ref{den}, we plot the density
distribution of member stars which are also plotted in Figure \ref{xy}. Figure
\ref{den} shows more clearly that most of member stars lie in the region with a
radii of 10 arcmin.

\section{Conclusions\label{conc}}
In this paper, using BATC 13 bands photometric data, we develop a SED method to
determine membership and fundamental parameters of a star cluster
simultaneously. Membership probabilities of 750 stars with limiting magnitude
15.0 in BATC $c$ band are derived for open cluster M48, 323 stars with
membership probabilities higher than 30\% are considered as member stars of M48.
Comparing with the membership determinations of 229 common stars taken from the
parametric \citep{wu02b} or non-parametric \citep{ba05} proper-motion based
methods, we get 80\% agreement with them. The present SED method can investigate
membership for stars with fainter magnitude than that of proper-motion based
methods (about 2 magnitude deeper in the case of M48) and membership
probabilities of 521 stars in the field of M48 are derived at the first time. At
the same time, the fundamental parameters of M48 are also derived and consist
with the previous determinations \citep{ri04, wu05, ba05}.

\acknowledgments

This research has made use of the Astrophysical Integrated Research Environment
(AIRE) which is operated by the Center for Astrophysics, Tsinghua University. We
also thank an anonymous referee for a number of suggestions that improved the
rigor and clarity of the paper. This work has been supported in part by the
Chinese National Science Foundation, No. 10473012, 10573020, and 10373020.

\newpage

\begin{deluxetable}{crrrrrrrrrrrrrcccc}
\tabletypesize{\scriptsize} \tablecolumns{18} \tablewidth{0pt} \rotate
\tablecaption{Member Stars Determined by both the present SED method and the
proper-motion based methods.\label{tb1}}
\tablehead{\colhead{No.}&\colhead{$c$}&\colhead{$d$}&\colhead{$e$}&\colhead{$f$}
&\colhead{$g$}&\colhead{$h$}&\colhead{$i$}&\colhead{$j$}&\colhead{$k$}&\colhead{
$m$}&\colhead{$n$}&\colhead{$o$}&\colhead{$p$}&\colhead{P$_{SED}$}&\colhead{
P$_W$}&\colhead{ P$_B$}&\colhead{IDs}} \startdata
    68&11.271&11.182&11.204&11.221&11.270&11.341&11.434&11.345&11.507&11.533&11.581&11.593&11.576&0.99&0.96&0.85& 284\\
& 0.0169& 0.0075& 0.0324& 0.0300& 0.0113& 0.0085& 0.0188& 0.0111& 0.0169& 0.0090& 0.0099& 0.0113& 0.0135&&&&\\
    99&12.789&12.644&12.577&12.533&12.486&12.542&12.511&12.438&12.464&12.501&12.492&12.520&12.474&0.98&0.94&0.73& 459\\
& 0.0205& 0.0106& 0.0326& 0.0300& 0.0113& 0.0095& 0.0190& 0.0116& 0.0175& 0.0103& 0.0115& 0.0114& 0.0178&&&&\\
   189&11.705&11.639&11.633&11.609&11.623&11.682&11.737&11.694&11.743&11.812&11.846&11.872&11.852&1.00&0.95&0.85& 268\\
& 0.0181& 0.0080& 0.0324& 0.0298& 0.0114& 0.0086& 0.0187& 0.0114& 0.0166& 0.0089& 0.0101& 0.0114& 0.0141&&&&\\
   223&10.338&10.354&10.418&10.413&10.506&10.599&10.721&10.658&10.793&10.853&10.908&10.917&10.944&0.74&0.96&0.85& 266\\
& 0.0167& 0.0076& 0.0324& 0.0297& 0.0111& 0.0082& 0.0186& 0.0101& 0.0165& 0.0087& 0.0097& 0.0102& 0.0116&&&&\\
\enddata
\end{deluxetable}
 \begin{deluxetable}{crrrrrrrrrrrrrcccc}
\tabletypesize{\scriptsize} \tablecolumns{18} \tablewidth{0pt} \rotate
\tablecaption{Member Stars Determined only by the present SED
method.\label{tb2}}
\tablehead{\colhead{No.}&\colhead{$c$}&\colhead{$d$}&\colhead{$e$}&\colhead{$f$}
&\colhead{$g$}&\colhead{$h$}&\colhead{$i$}&\colhead{$j$}&\colhead{$k$}&\colhead{
$m$}&\colhead{$n$}&\colhead{$o$}&\colhead{$p$}&\colhead{P$_{SED}$}&\colhead{
P$_W$}&\colhead{ P$_B$}&\colhead{IDs}} \startdata
     2&14.316&13.881&13.756&13.627&13.450&13.479&13.407&13.305&13.365&13.321&13.310&13.295&13.270&0.91&0.00&0.00&   0\\
& 0.0458& 0.0108& 0.0345& 0.0312& 0.0204& 0.0137& 0.0212& 0.0142& 0.0237& 0.0113& 0.0124& 0.0140& 0.0218&&&&\\
     4&13.168&12.972&12.927&12.903&99.999&99.999&12.741&12.620&12.659&12.649&99.999&12.679&12.584&0.97&0.01&0.00& 461\\
& 0.0310& 0.0141& 0.0352& 0.0313&&& 0.0193& 0.0178& 0.0200& 0.0124&& 0.0170& 0.0306&&&&\\
     6&14.751&14.547&14.413&14.282&14.138&14.138&14.038&13.930&13.982&13.935&13.912&13.887&13.872&0.75&0.00&0.00&   0\\
& 0.0472& 0.0108& 0.0342& 0.0311& 0.0177& 0.0124& 0.0203& 0.0190& 0.0199& 0.0139& 0.0151& 0.0168& 0.0238&&&&\\
    17&14.934&14.845&14.657&14.600&14.380&14.383&14.268&14.130&14.115&14.165&14.151&14.102&14.064&0.33&0.00&0.00&   0\\
& 0.0530& 0.0439& 0.0409& 0.0348& 0.0274& 0.0267& 0.0230& 0.0322& 0.0268& 0.0206& 0.0270& 0.0259& 0.0338&&&&\\
\enddata
\end{deluxetable}

\begin{deluxetable}{crrrrrrrrrrrrrcccc}
\tabletypesize{\scriptsize} \tablecolumns{18} \tablewidth{0pt} \rotate
\tablecaption{Field Stars Determined by the present SED method but considered as
member stars by the proper-motion based methods.\label{tb3}}
\tablehead{\colhead{No.}&\colhead{$c$}&\colhead{$d$}&\colhead{$e$}&\colhead{$f$
}&\colhead{$g$}&\colhead{$h$}&\colhead{$i$}&\colhead{$j$}&\colhead{$k$}&\colhead{
$m$}&\colhead{$n$}&\colhead{$o$}&\colhead{$p$}&\colhead{P$_{SED}$}&\colhead{
P$_W$}&\colhead{P$_B$}&\colhead{IDs}} \startdata
     3&11.528&11.210&11.090&11.013&&&10.674&10.595&10.547&10.534&&10.523&10.496&0.00&0.87&0.88& 293\\
& 0.0174& 0.0072& 0.0328& 0.0301& & & 0.0188& 0.0154& 0.0167& 0.0090& & 0.0096& 0.0114&&&&\\
   406& 9.529& 8.903& 9.292& 8.120& 7.964& 7.894& 7.324& 7.601& 6.810& 7.837& 7.285& 7.293& 7.245&0.00&0.88&0.84& 234\\
& 0.0164& 0.0071& 0.1442& 0.1079& 0.0120& 0.0344& 0.0549& 0.0444& 0.0615& 0.0442& 0.0522& 0.0098& 0.0103&&&&\\
   474&10.991&10.365&10.008& 9.826& 9.468& 9.417& 9.208& 9.084& 8.993& 8.975& 8.868& 8.816& 8.779&0.00&0.07&0.84& 424\\
& 0.0170& 0.0071& 0.0323& 0.0297& 0.0111& 0.0084& 0.0189& 0.0114& 0.0164& 0.0085& 0.0091& 0.0096& 0.0108&&&&\\
   498&12.145&12.018&11.983&11.942&11.881&11.916&11.884&11.837&11.845&11.895&11.893&11.915&11.899&0.00&0.95&0.82& 212\\
& 0.0190& 0.0081& 0.0325& 0.0299& 0.0114& 0.0086& 0.0187& 0.0100& 0.0167& 0.0095& 0.0104& 0.0107& 0.0143&&&&\\
\enddata
\end{deluxetable}

\end{document}